\documentclass[aps,prl,twocolumn,superscriptaddress]{revtex4}
\usepackage{amsmath}
\usepackage{amssymb}
\usepackage{graphicx}
\usepackage{dcolumn}
\usepackage{bm}
\makeatletter
\def\captionof#1#2{{\def\@captype{#1}#2}}
\makeatother

\begin{document}

\title{
Critical quench dynamics in confined systems
}
\author{Mario Collura}
\affiliation{Institut Jean Lamour, dpt. P2M, Groupe de Physique Statistique, Nancy-Universit\'e CNRS,
B.P. 70239, F-54506 Vandoeuvre les Nancy Cedex, France}
\affiliation{Institut f\"ur Theoretische Physik, Universit\"at Leipzig, Postfach 100\,920, 04009 Leipzig, Germany}

\author{Dragi Karevski}
\affiliation{Institut Jean Lamour, dpt. P2M, Groupe de Physique Statistique, Nancy-Universit\'e CNRS,
B.P. 70239, F-54506 Vandoeuvre les Nancy Cedex, France}


\begin{abstract}
We analyze the coherent quantum evolution of a many-particle system after slowly sweeping a power-law confining potential. The amplitude of the confining potential is varied in time along a power-law ramp such that the many-particle system finally reaches or crosses a critical point. Under this protocol we derive  
general scaling laws for the density of excitations created during the non-adiabatic sweep of the confining potential. It is found that the mean excitation density follows an algebraic law as a function of the sweeping rate with an exponent that depends on the space-time properties of the potential. 
We confirm our scaling laws by first order adiabatic calculation and exact results on the Ising quantum chain with a varying transverse field.
\end{abstract}

\pacs{Valid PACS appear here}
\maketitle

Recent beautiful experiments with ultracold atomic gases have revitalized the studies 
of nonequilibrium aspects in strongly correlated quantum systems \cite{Exp1}. The main peculiarities of the dynamics of ultracold atoms are low dissipation rate and phase coherence over very long times \cite{Exp1} such that the dynamics is very well described by the usual quantum unitary evolution of closed systems.      
An important problem that has received much attention is the situation where the parameters of the quantum many-body system are varied in time such that it reaches or crosses a quantum critical point \cite{Sachdev}. The divergence of the intrinsic relaxation time close to the critical point induces non-adiabatic evolution no matter how slow is the rate of change of the Hamiltonian. If the system is initially in its ground state non-adiabatic transitions toward excited states lead to the presence of topological defects in the final state \cite{ZuDoZo05,Da05,Dz05,Po05,BaPo08}. For example, driving a quantum system from a paramagnetic to a ferromagnetic phase through a critical point generates a final state given by a superposition over excited states carrying finite ferromagnetic domains separated by kinks or domain walls. 
For a slow driving rate the density of defects is a universal scaling function of the driving rate as in the classical Kibble-Zurek (KZ) mechanism  \cite{Kibble}. 
This may be of importance in the context of adiabatic quantum computation \cite{Farhi}
 where adiabatic evolution is proposed to transfer the system from an initial state to a computational nontrivial state. If one is forced to cross a critical point in order to generate the nontrivial state, inevitably the crossing will result in the generation of excitations (defects).   The optimal time ramp needed to drive the system through the critical point has to balance the unavoidable generation of defects and the time needed to cross the critical point \cite{BaPo08}.

\begin{figure}[t]
\includegraphics[width=0.43\textwidth]{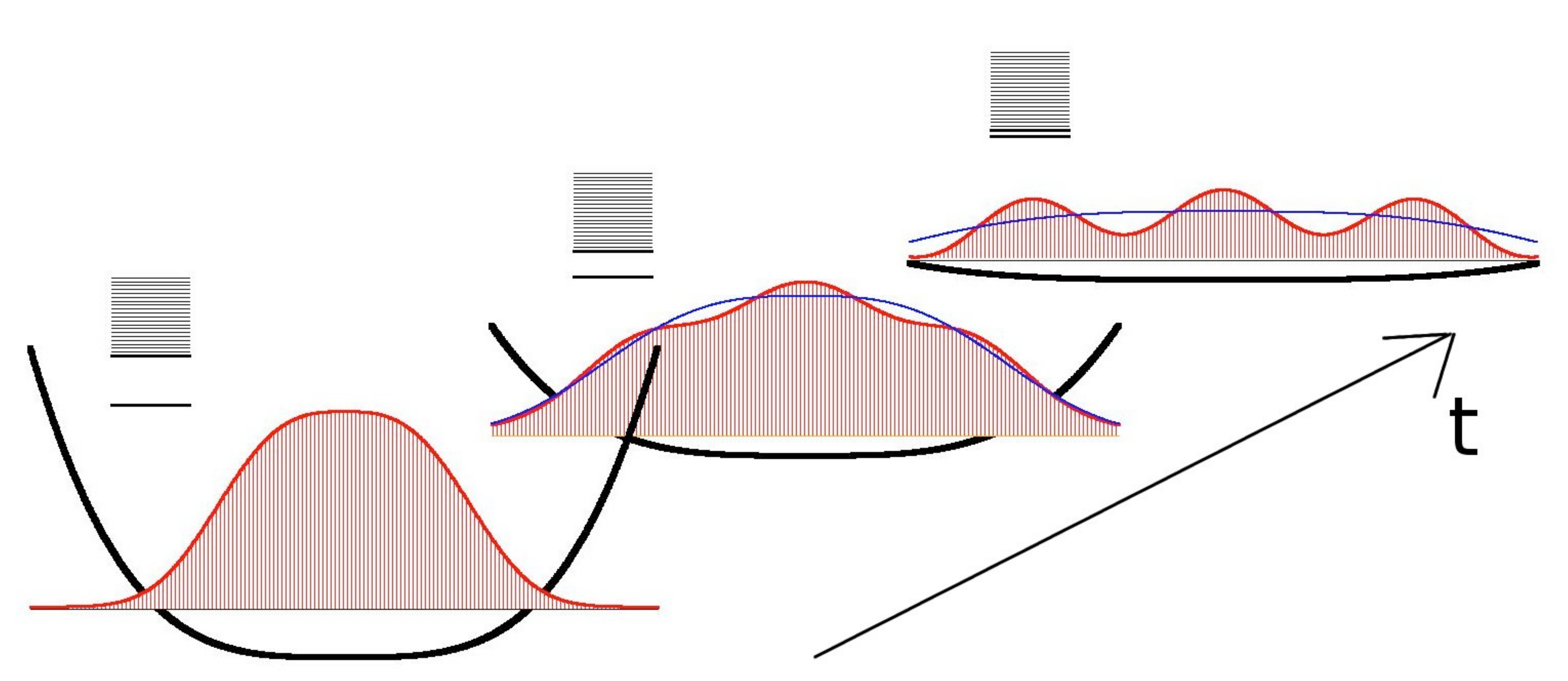}
\caption[width=\textwidth]{\label{paper_fig01} Sketch of the slow sweep of a power law confining potential. The state of the system (red) deviates from the adiabatic ground state (blue) as we approach the critical point associated to a vanishing potential. The small stripes represent the spectrum at different times.   }
\end{figure}

In many real physical situations the critical systems under study are immersed into an inhomogeneous field in one or several directions (for example the gravity field) whose main effect is to smooth out the critical singularities \cite{PlKaTu07,ZuDo08,CoKaTu09}. This is particularly relevant in the context of ultra-cold atoms where parabolic trapping potentials are used to confine the atomic cloud.
Removing or loading smoothly in time such a power-law trap will lead to a final state carrying a nontrivial density of defects, depending on the shape of the trap, as soon as we are getting closer to a critical point. 
In this letter we analyze the coherent generation of defects in many-body quantum systems after slowly sweeping a power-law confining potential close to a critical point, see figure \ref{paper_fig01}. 
We develop a general scaling scenario from which we derive the scaling behavior of local (such as the local energy density or order parameter) and global quantities (such as the density of defects). We derive in particular the optimal non-linear time ramp minimizing the generation of defects for a given total sweeping time. The scaling theory is tested on the Ising quantum chain with an inhomogeneous transverse field playing the role of the confining potential.

{\sc  Scaling theory}
Consider a system which has at zero temperature a quantum critical point, controlled by a scalar field $h$, separating a symmetric phase from a broken-symmetric one. Assume that close enough to the critical point the control parameter $h$ deviates in one direction from its critical value $h_c$ with a power law
\begin{equation}\label{perturb}
\delta(x,t)\equiv h(x,t)-h_c \simeq g(t) |x|^\omega\; .
\end{equation}
The amplitude $g(t)$ of the deviation is driven externally and varies in time as $g(t)=v|t|^\alpha {\rm sgn}(t)$ 
where, without loss of generality, the rate amplitude $v$ is positive.
The space and time exponents $\omega$ and $\alpha$ are both positive. 
Notice that within this parametrization the quench dynamics connects the two phases ($\delta>0$ and $\delta<0$) by crossing the homogeneous critical point  at $t=0$.  

The inhomogeneous perturbation  (\ref{perturb}) introduces a crossover region in space-time $(x,t)$ around the critical locus $(0,0)$
with characteristic length-scale $\ell$ and time-scale $\tau$.
Indeed, starting in the far past in the ground state $|0(g(-\infty))\rangle$ the system evolves adiabatically  toward the instantaneous ground state  $|0(g(t))\rangle$ as far as it is protected by a large energy gap $\Delta(t)$.
In the instantaneous state $|0(g(t))\rangle$, the spacial power-law variation of the control parameter $\delta$ introduces a typical length scale $\ell(t)$ around the spatial critical locus (here at $x=0$) \cite{PlKaTu07}.  The typical length can be obtained self-consistently from the correlation length relation $\ell(t)\sim \delta(\ell,t)^{-\nu}$. One finds $\ell(t)\sim |g(t)|^{-\nu/(1+\nu\omega)}$ diverging for a vanishing deviation amplitude $g$.
Here $g(t)$ plays the role of an effective deviation to the critical point with  effective exponent $\nu_g=1/y_g=\nu/(1+\nu\omega)$.
 As time runs toward zero, the gap closes and the relaxation time $\propto 1/\Delta$ grows up to the point where the adiabatic approximation breaks down completely. Getting closer and closer to the critical point the dynamics switches to a sudden regime. For larger times, after the critical point has been crossed, one recovers again the nearly adiabatic regime. This is the Kibble-Zurek (KZ) scenario.
The typical time-scale $\tau$ around the critical locus where this break-down occurs can be deduced from the self-consistent relation $\tau\sim \ell(\tau)^z$, with $z$ the dynamical exponent. It leads to
$\tau\sim\ell^z\sim v^{- z/y_v}$ where $y_v=y_g+z\alpha=(1+\nu(\omega+z\alpha))/\nu$
is the RG dimension of the perturbation field, such that under rescaling by a factor $b$ the amplitude $v$ transforms as $v'=b^{y_{v}}v$. Since $\omega$ and $\alpha$ are positive, the perturbation is always relevant ($y_v>0$). 

{\it Local and bulk quantities} The profile of a local quantity $\varphi(x,t,v)$, like the local order parameter or the energy density,  with scaling dimension $x_{\varphi}$ transforms under rescaling as
$
\varphi(x,t,v) = b^{-x_{\varphi}}\varphi(xb^{-1},tb^{-z},vb^{y_{v}}).
$
Taking $b=v^{-1/y_{v}}\propto\ell\propto\tau^{1/z}$, one obtains
\begin{eqnarray}\label{spacetime_profile}
\varphi(x,t,v) & = & v^{x_{\varphi}/y_{v}}\Phi(xv^{1/y_{v}},tv^{z/y_{v}})\; .
\end{eqnarray}
The prefactor exhibits the trap-size scaling $\varphi\sim \ell^{-x_\varphi}$ associated to a finite size system with $\ell\sim v^{-1/y_{v}}$ \cite{PlKaTu07}. The unknown scaling function $\Phi$ encodes the propagation 
in space of the critical phase (close to $x=0$) into the initial one. See \cite{KibVol} for detailed studies on that issue.
A bulk quantity or a spatial averaged profile $\overline{\varphi}(t,v)$ satisfies
\begin{equation}\label{time_profile}
\overline{\varphi}(t,v)=v^{x_{\varphi}/y_{v}}\overline{\Phi}\left(t v^{z/y_{v}}\right)\sim
\tau^{-x_{\varphi}/z}\overline{\Phi}\left(\frac{t}{\tau}\right)\; .
\end{equation}
As an example, the spatial averaged energy density, with scaling dimension $x_e=d+z$, should behave after the quench to the critical point as $e\sim v^{(d+z)/y_v}$.

{\it Density of defects}
An estimate of the density of defects $n$ generated by crossing the critical point with the nonlinear ramp $g(t)$ is obtained from the length scale identified above through the relation $n\sim \ell^{-d}\sim v^{d/y_v}$. 
To be more precise, equating the relaxation time $\tau_0/\Delta(t)$ with the typical time-scale $\Delta(t)/|\dot{\Delta}(t)|$ at which the Hamiltonian is varied and assuming that the gap scales as $\Delta(t)\simeq \Omega_0 |g|^{z/y_g}$ one finds for the typical (Kibble-Zurek) time-scale 
$
\tau_{KZ}\sim \left(\frac{\tau_{0}}{\Omega_0}\frac{z\alpha}{y_{g}}\right)^{y_{g}/y_{v}}v^{-z/y_{v}}
$.
From the relation $n\sim [\Delta(\tau_{KZ})]^{d/z}$ we get the behaviour
\begin{equation}\label{excitation_density_KZM}
n\sim \left(\frac{\tau_{0}}{\Omega_0}\frac{z\alpha}{y_{g}}\right)^{d\alpha/y_{v}}v^{d/y_{v}}=(z\gamma\delta)^{\frac{d\gamma}{1+z\gamma}}
\end{equation}
with $\delta=\frac{\tau_{0}}{\Omega_0}v^{1/\alpha}\sim1/T$ ($T$ defining the temporal window of the quench protocol) and $\gamma=\alpha\nu_g=\alpha/y_{g}$. 
As expected the density of defects $n$ is smaller for larger values of the protocol time window $T$ necessary to reach a final value $g_f$ from the initial $g_0$ one.
If one wants to minimize the generation of defects, the switching of the trap should be as slow as possible. However this can lead to extremely long protocol times $T$ and be counterproductive, as for example in quantum computational issues where one looks for a compromise between the production of excited states  and short computational times.
To achieve this compromise one may look for the optimal power-law time ramp protocol that minimizes the defect density $n$ at a fixed duration $T$.  
Extremizing  (\ref{excitation_density_KZM}) with respect to $\gamma=\alpha/y_{g}$ for a given $\delta\sim 1/T$ one finds 
\begin{equation}
\gamma_{opt} = \frac{1}{z}\mathrm{W}\left(\frac{1}{e\delta}\right),
\end{equation}
where $\mathrm{W}(x)$ is the Lambert W function defined through  $x=f(W) = We^W$. For a given trap shape (space exponent $\omega$ fixed)  the optimum time exponent takes the value $\alpha_{opt}=
\gamma_{opt} /\nu_g=\gamma_{opt}(1+\nu\omega)/\nu$. The result of \cite{BaPo08} is recovered with $\omega=0$ by expanding the Lambert function. We see that loading a trap potential  changes significantly the homogeneous optimal protocol by increasing the exponent $\alpha$ by a factor $(1+\nu\omega)$ (close to the critical point one has to be slower than in the homogeneous case).

{\sc Dynamics in the Ising chain}
The Ising quantum chain is a standard theoretical laboratory for issues related to quantum phase transitions \cite{Sachdev} and it has been already studied in various sudden quench dynamics contexts \cite{Qu1}. In this model, an inhomogeneity analogous to the one arising from a trapping potential in particle systems can be achieved by considering a time-dependent inhomogeneous transverse field. The Hamiltonian is
\begin{equation}\label{H(t)_pauli}
\mathcal{H}(g(t)) = -\frac{1}{2}\sum_{n=1}^{L-1}\sigma^{x}_{n}\sigma^{x}_{n+1} - \frac{1}{2}\sum_{n=1}^{L}h_n(g(t))\sigma^{z}_{n},
\end{equation}
where $h_n(g(t))=1+g(t)n^{\omega}$, with $g(t)=v|t|^{\alpha}\mathrm{sgn}(t)$. 
Notice that the spatial critical locus has been set at the left boundary of the chain. For the critical unperturbed system, $g=0$, the dynamical exponent $z=1$ and the correlation length exponent $\nu=1$. 

If the rate of change of the Hamiltonian is slow enough one can use a nearly adiabatic approximation in order to describe the actual state 
$|\Psi(t)\rangle$
generated from the initial ground state $|0(t_0)\rangle$
by the unitary time evolution.
Introducing the instantaneous eigenbasis $\{|k(t)\rangle\}$, such that $\mathcal{H}(g(t))|k(t)\rangle=E_k(t)|k(t)\rangle$, one obtains from standard perturbation theory  the ``one-jump'' expansion 
$
|\Psi(t)\rangle \simeq  \mathrm{e}^{-i\vartheta_0(t_0,t)} (|0(t)\rangle
+ \sum_{k\neq0}    a_k(t_0,t)     |k(t)\rangle), 
\label{adiabatic_expansion}
$
where $\vartheta_k(t_0,t)=\int_{t_0}^t ds \; E_k $ is the dynamical phase. The transition amplitudes $a_k(t_0,t)$ are given by 
\begin{equation}\label{adiabatic_amplitude}
a_{k}(t_0,t)  =  \int_{g(t_0)}^{g(t)}dg \frac{ \langle k(g)|\partial_{g}\mathcal{H}(g)|0(g)\rangle}{\delta\omega_{k0}(g)} \mathrm{e}^{-i\vartheta_{k0}(g,g(t))} \; 
\end{equation}
with $\delta\omega_{k0}(t)= E_{k}(t)-E_{0}(t)$ the Bohr frequencies and 
$\vartheta_{k0}=\vartheta_k-\vartheta_0$ a dynamical phase factor. 
Under a Jordan-Wigner fermionization followed by a canonical Bogoliubov mapping the Ising Hamiltonian (\ref{H(t)_pauli}) is diagonalized in terms of the
Dirac fermionic algebra $\{\eta^{\dag}_{p}(g), \eta_{q}(g)\}=\delta_{pq}$ and reads
$
\mathcal{H}(g(t))=\sum_p\epsilon_{p}(g)\left[\eta^{\dag}_{p}(g)\eta_{p}(g)-1/2\right]
$
where $\epsilon_{p}(g)$ are the instantaneous one-particle excitation energies. 
The perturbation $\partial_{g}\mathcal{H}(g)$ in (\ref{adiabatic_amplitude}) takes the form
\begin{equation}\label{H_derivative}
\partial_{g}\mathcal{H}(g) = \frac{1}{2}\sum_{p,q}X^{\omega}_{pq}(g)[\eta^{\dag}_{p}(g)+\eta_{p}(g)][\eta^\dag_{q}(g)-\eta_{q}(g)],
\end{equation}
with $X^{\omega}_{pq}(g)=\sum_n\phi_p(n,g)n^\omega \psi_q(n,g)$  where the $\phi$ and $\psi$ are the Bogoliubov coefficients. This perturbation induces transitions from the ground state  to the two-particles states $|pq(g)\rangle=\eta^{\dag}_{q}(g)\eta^{\dag}_{p}(g)|0(g)\rangle$ only. These fermionic particles are the basic topological defects in the Ising chain.
At time $t$, the density of defects in the mode $q$ is given by the instantaneous population 
$n_q(t)=\langle \Psi(t)| \eta_q^\dag(g) \eta_q(g)|\Psi(t)\rangle \simeq 4\sum_p |a_{pq}(t_0,t)|^2$.

{\it  Linear spatial perturbation}
For a linear spatial modulation, that is at $\omega=1$, we have an explicit solution in the thermodynamical limit $L\rightarrow \infty$: the Bogoliubov coefficients are given by the wave-functions $\chi_p$ (up to normalization) of a simple harmonic oscillator  \cite{CoKaTu09}.
The  matrix elements $\langle pq(g)|\partial_{g}\mathcal{H}(g)|0(g)\rangle$ in (\ref{adiabatic_amplitude}) are then
proportional to the position matrix elements of the harmonic oscillator $(\chi_p,u\chi_q)$, such that the only non vanishing transition amplitudes $a_{pq}(t_0,t)$ are those with $p=q\pm 1$.

Plugging the exact solution into (\ref{adiabatic_amplitude}) one obtains a closed expression for the amplitudes $a_{pq}$. However,
contrary to the spatial homogeneous case ($\omega=0$) where the integral (\ref{adiabatic_amplitude}) converges at the critical value $g=0$, here the linear spatial inhomogeneity modifies the dependence on $g$ of the integrant to a $g^{-1}$ behavior leading to a logarithmic divergence at $g=0$. This divergence is caused by the square root dependence on $|g|$ of the excitation spectrum $\epsilon_p= |g|^{1/2}\sqrt{4p+1+\mathrm{sign}(g)}$ with $p=0,1,2,...$ \cite{CoKaTu09}. Consequently, the first order adiabatic expansion breaks down at the critical point $g=0$ (at $t=0$). 
Nevertheless, for quenches that do not cross the critical point (the starting and the ending point of $g$ are on the same side of the critical locus) one can still use (\ref{adiabatic_amplitude}) giving
\begin{equation}
\label{apq}
|a_{pq}(t_0,t)|^2=|\frac{G_{pq}}{4\Omega_{pq}} {A}_{\rho_{pq}}\left(|g_{0}|,|g(t)|\right)|^2
\end{equation}
with $G_{pq}= \sqrt{{2(p+q)+1+{\rm sgn}(g)}}\left[\delta_{p\,q-1}-\delta_{p\,q+1}\right]$
$\Omega_{pq} = |g|^{-1/2} \delta\omega_{pq,0}(g)$ and
$\rho_{pq}  =  -2\Omega_{pq}\frac{v^{-1/\alpha}}{\alpha+2}{\rm sgn}(g)$.
The function 
\begin{equation}\label{A_Ei_function}
{A}_{\rho}(x,y) = \frac{2\alpha}{2+\alpha}\left[\mathrm{E_{1}}\left(i\rho x^{\frac{2+\alpha}{2\alpha}}\right)-\mathrm{E_{1}}\left(i\rho y^{\frac{2+\alpha}{2\alpha}}\right)\right]
\end{equation}
is expressed in terms of the exponential integral  $\mathrm{E_{1}}(z)=\int_{z}^{\infty}dt\,t^{-1}\mathrm{e}^{-t}$ for $\left|\mathrm{Arg}(z)\right|<\pi$.

In order to compare the analytical result (\ref{apq}) with our scaling predictions consider first the case where 
the quench starts far away from the critical locus, $|t_0|\gg 1$. In that case $\mathrm{E_{1}}\left(i\rho |g_0|^{\frac{2+\alpha}{2\alpha}}\right)\simeq 0$ and the function $A(x,y)$ is dominated by $E_1(i\rho |g(t)|^{\frac{2+\alpha}{2\alpha}}) $ leading to the expected scaling behavior 
$
n_q(t,v)=f(v|t|^{2+\alpha})
$ 
with $y_v=2+\alpha$, see (\ref{time_profile}).
On the contrary if the initial trap amplitude $g_0$ is small enough one observes the non-homogeneous behavior 
$
n_q(g_{0};t,v)\sim f_0(g_0,v) +  f_1(|t|v^{1/y_{v}})
$. The expected scaling behavior (\ref{time_profile}) is broken by the presence of the boundary term $f_0$ which accounts for the high correlations in the initial ground state $|0(g_0)\rangle$ (since for $|g_0|\ll 1$ the system is nearly critical).

For a quench at or crossing the critical point the situation is more complicated since, as stated before, the (un-normalised) perturbation formula (\ref{adiabatic_amplitude}) leads to a divergence at $t=0$. However, for a finite size chain the energy gap $\delta\omega_{k0}$ stays finite at the critical point which wash out the critical divergences and one can perform a finite size scaling study. 

\begin{figure*}[t]
\includegraphics[width=0.32\textwidth]{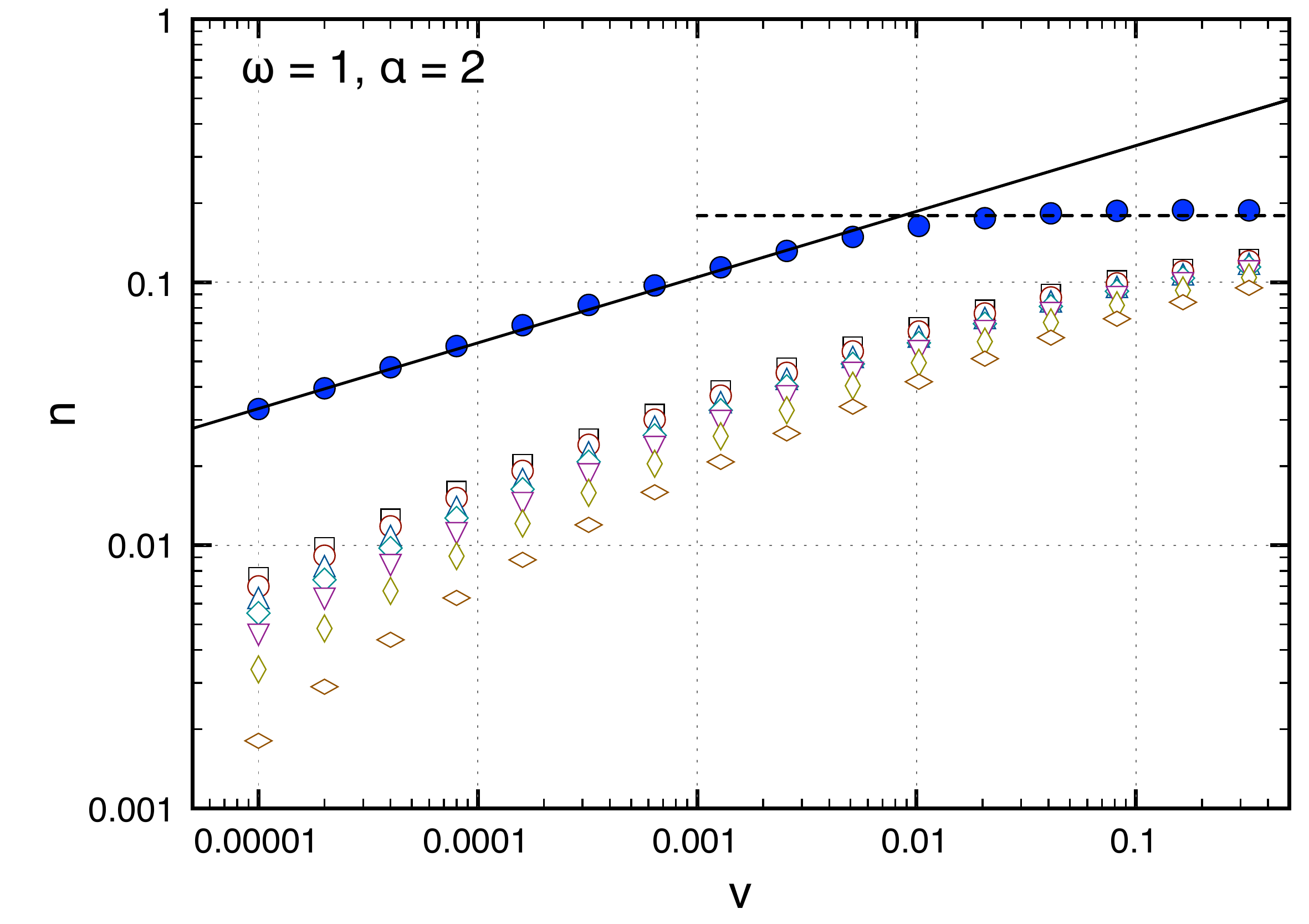}
\includegraphics[width=0.32\textwidth]{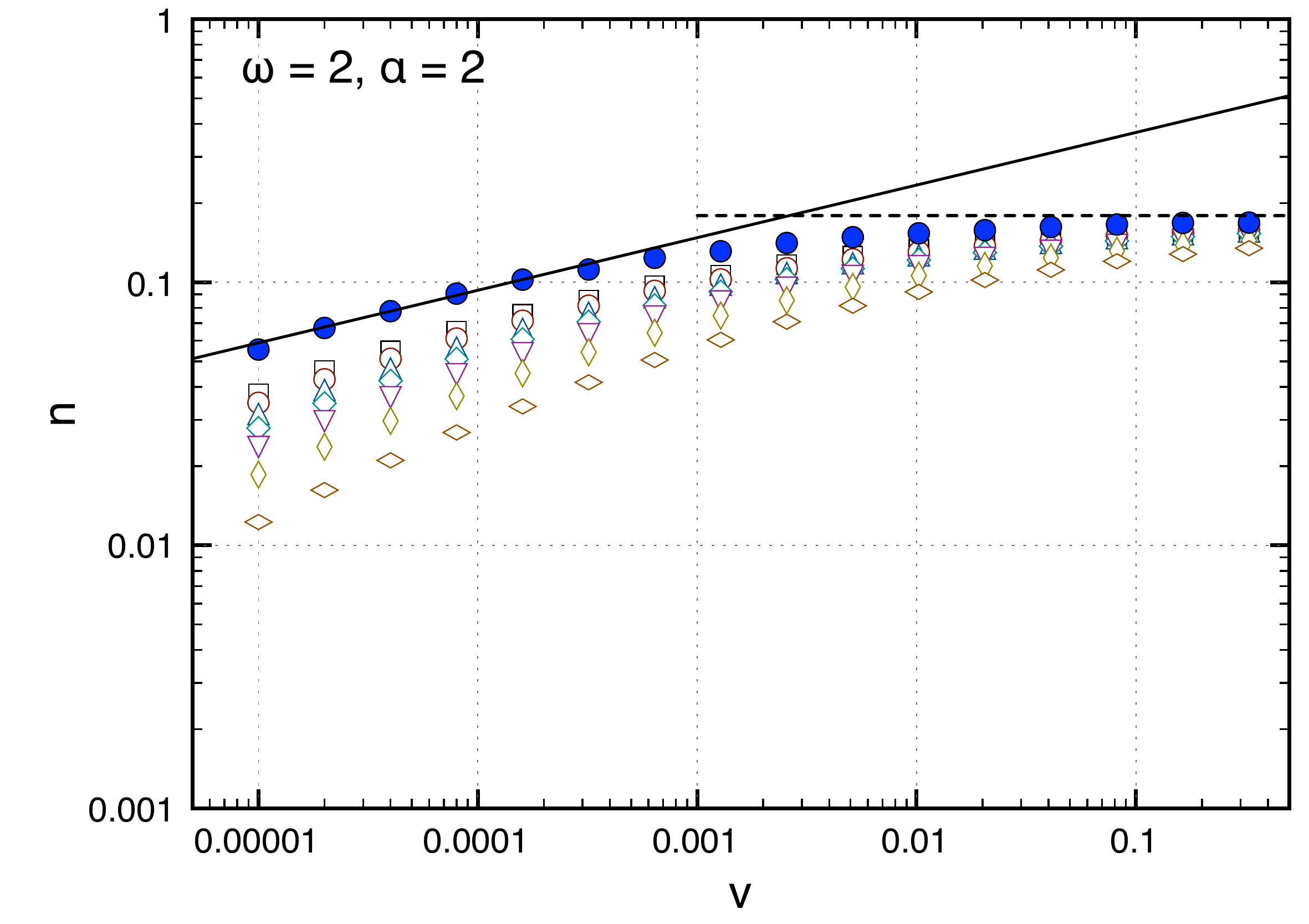}
\includegraphics[width=0.32\textwidth]{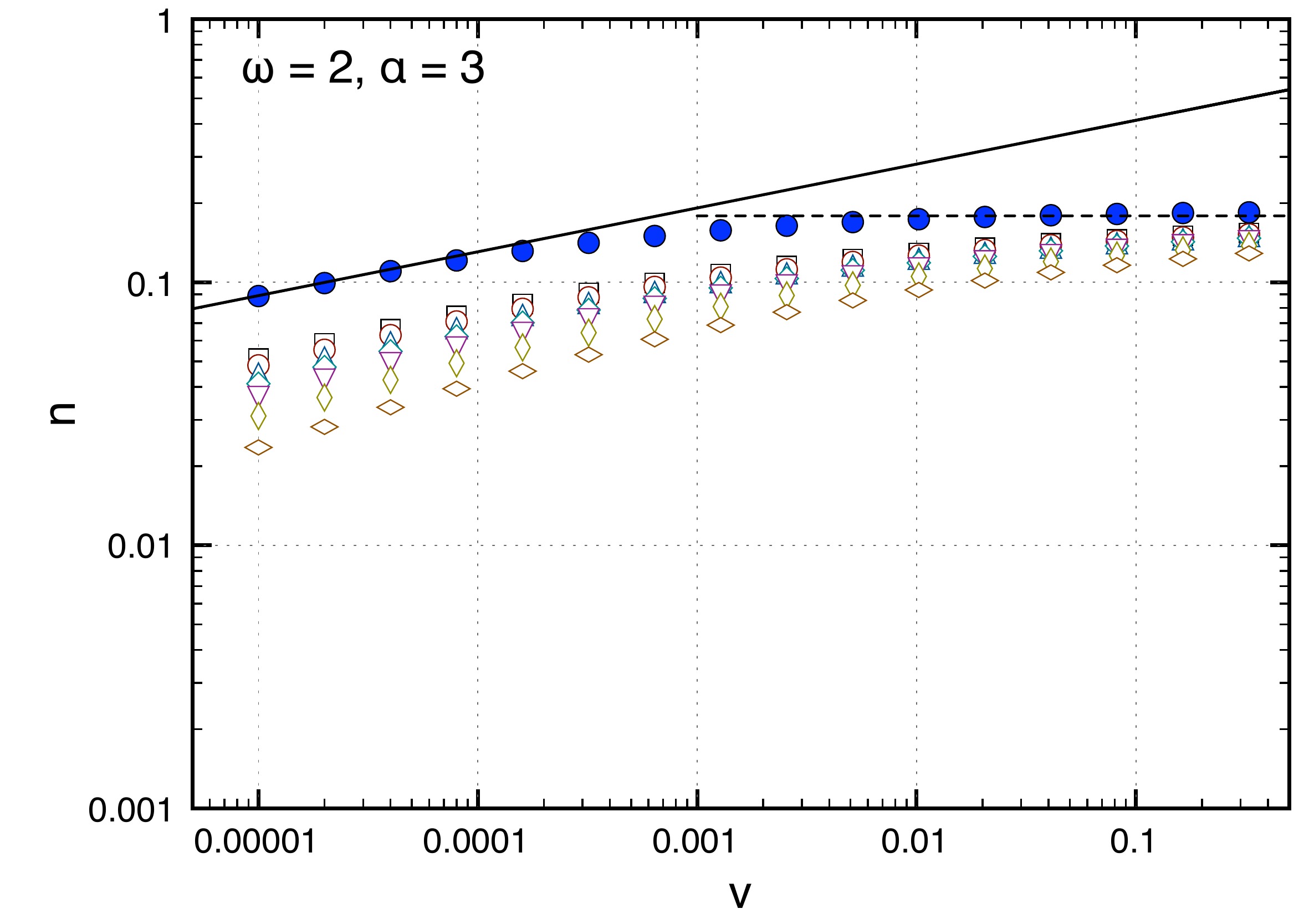}\\
\caption[width=\textwidth]{\label{paper_fig2} Density of defects $n$  versus the quench parameter $v$ for a critical quench. The amplitude changes from $g=1$ to $g=0$. Empty symbols correspond to different system sizes ($L = 96$ to $L=256$ from bottom to top). The extrapolated data (filled circles) show, in the adiabatic limit $v\ll 1$, a perfect agreement with the scaling prediction: $n\sim v^{1/4}$ for $\omega=1,\,\alpha=2$; $n\sim v^{1/5}$ for $\omega=2,\,\alpha=2$; $n\sim v^{1/6}$ for $\omega=2,\,\alpha=3$ (straight lines). The dashed lines give the sudden-quench value $n_{sq}\approx 0.179$ evaluated on a system with $L=1024$ spins. }
\end{figure*}

{\it Finite size scaling analysis for general $\omega$}  
We start far away in the disordered phase ($g=1$)  and drive the system to the critical point ($g=0$). The density of defects is obtained numerically by exact diagonalisation of finite chains with up to 256 spins. The finite size data are then extrapolated to the infinite size limit. The results are reported on figure \ref{paper_fig2} for different spatial and temporal exponents $\omega$ and $\alpha$. 
First of all we observe that the finite size values are always smaller than the extrapolated ones: indeed, on the finite system the gap does not vanish
even close to the critical locus and consequently the generation of defects is smaller than the expected one in the  thermodynamical limit. 
At small $v$ the extrapolated data are in perfect agreement with the scaling prediction (\ref{excitation_density_KZM}) which is represented by the full line. 
The saturation at large $v$ of the defects production, independently of the quench protocol ($\alpha$ and $\omega$ values), is due to the fact that for very fast quenches the only relevant parameters are the initial and final amplitudes $g$.  If the initial amplitude is very high in modulus, the correlation length is very small (of order one) and there is no big differences for different values of $\omega$. One expects the same defect production as in the case of a sudden quench to the critical point of a completely disordered initial state (which is given by the horizontal dashed line). 
{A similar behavior was reported in \cite{KibVol} where for fast enough inhomogeneous quenches one recovers the homogeneous defect production (corresponding here to the homogeneous sudden quench saturation at large $v$) while  for slow enough inhomogeneous quenches (here small $v$) the defect production is significantly lowered.

{\sc Conclusion} 
We have obtained the scaling properties of the density of defects generated during the loading or unloading of a power-law trap. The analysis is based on the identification of a KZ-time scale which depends on the universal properties of the critical point as well as on the exponents characterizing the temporal ramp and the spatial trap. 
We found that the optimal nonlinear passage through the quantum critical point, specified by an optimal time ramping exponent $\alpha_{opt}$, is strongly affected by the presence of the trapping potential.
The analyses made on the quantum Ising chain revealed  quite strong finite size corrections, as seen on figure \ref{paper_fig2}, to the expected scaling prediction (\ref{excitation_density_KZM}) for the density of defects.    
A relevant extension of this work is the study of the influence of a finite temperature on the scaling properties \cite{patane}. This is currently under investigation.

This work is supported by ANR-09-BLAN-0098-01. 


\end{document}